\documentclass[fleqn,usenatbib]{mnras}

\usepackage{newtxtext,newtxmath}

\usepackage[T1]{fontenc}
\usepackage{ae,aecompl}

\usepackage{graphicx}	
\usepackage{amsmath}	

\usepackage[switch]{lineno}

\title[Strong constraints on spherical galactic models]{New strong constraints on the central behaviour of spherical galactic models -- No NFW cusp}

\author[M. Roncadelli and G. Galanti]{
Marco Roncadelli$^{1,2}$\thanks{E-mail: marco.roncadelli@pv.infn.it (MR)}
and Giorgio Galanti$^{3}$\thanks{E-mail: gam.galanti@gmail.com (GG)}
\\
$^{1}$INFN, Sezione di Pavia, Via A. Bassi 6, I -- 27100 Pavia, Italy\\
$^{2}$INAF, Osservatorio Astronomico di Brera, Via E. Bianchi 46, I -- 23807 Merate, Italy\\
$^{3}$INAF, Istituto di Astrofisica Spaziale e Fisica Cosmica di Milano, Via A. Corti 12, I -- 20133 Milano, Italy
}

\date{Accepted XXX. Received YYY; in original form ZZZ}

\pubyear{2021}

\begin{document}
\label{firstpage}
\pagerange{\pageref{firstpage}--\pageref{lastpage}}
\maketitle


\begin{abstract} 
We first stress that {\it any spherically symmetric} galactic model whose integrated mass profile $M (r) \to 0$ as $r \to 0$ is physically consistent close to the centre {\it only} provided that the circular velocity $v_c (r) \to 0$ and the gravitational field $g (r) \to 0$ as $r \to 0$. Next, we apply such a statement to a broad class of five-parameter spherical galactic models, which includes most of those used in astrophysics and cosmology. In particular, we discover that the Jaffe and Hernquist models can only be trusted for $r \gtrsim 0.2 \, R_e$, while the NFW model cannot describe the central region either of regular galaxy clusters or of pure dark matter halos, thereby failing to predict any central cusp. 
\end{abstract}



\section{Introduction} 
In spite of the steadily increasing precision of the N-body simulations of astronomical systems, spherical analytic models are still ubiquitous in astrophysics and cosmology, ranging from globular clusters to galaxies, and from regular galaxy clusters to dark matter halos. Examples are many. While the Plummer sphere model~\citep{plummer} provides a good description of stars in globular clusters, the Jaffe~\citep{jaffe} and Hernquist~\citep{hernquist} ones are routinely used to represent the stellar distribution in spheroidal elliptical galaxies and bulges. Further, the pseudo-isothermal profile (see e.g.~\citealt{pseudoisothermal}) correctly describes the intermediate part of dark matter halos of spiral galaxies while the Navarro-Frank-White (NFW) model~\citep{nfw} represents the galaxy distribution in regular clusters and 
provides a good analytic fit to the N-body simulations of pure dark matter halos within the standard $\Lambda$CDM context. 

The aim of the present paper is twofold. First, we consider a {\it generic spherically symmetric} galactic model whose integrated mass profile $M (r)$ -- defined by Eq. (\ref{a2}) below -- is such that $M (r) \to 0$ as $r \to 0$. And we stress that -- because of a certain statement -- the considered model is {\it physically consistent} near the centre {\it only} provided that two conditions are satisfied: the circular velocity $v_c (r) \to 0$ and the gravitational field $g (r) \to 0$ as $r \to 0$ (they are defined by Eqs. (\ref{a5}) and (\ref{a6}) below). Second, we apply such a statement to a broad class of five-parameter spherical galactic models, which includes most of those used in astrophysics and cosmology, in particular the three-parameter family of Dehnen profiles~\citep{dehnen,dehnengerhard,tremaine1994}, the Jaffe, the Hernquist and the pseudo-isothermal spherical models, as well as the NFW, the Plummer sphere, the modified Hubble~\citep{binneytremaine} and the perfect sphere~\citep{perfectsphere} profiles~\footnote{Observe that the \citet{dehnen} and \citet{tremaine1994} models coincide up to a simple rescaling of the radius.}. 

Surprisingly, nobody seems to have realized that the considered statement leads to new important results.

The paper is organized as follows. In Sect. 2 we detail and emphasize the above statement, while in Sect. 3 we describe our broad class of five-parameter spherical galactic models. Sect. 4 is devoted to the application of the statement in question to a few considered models from a purely mathematical point of view. We see that for some of them -- e.g. the Jaffe, Herquist and NFW profiles -- as the central distance decreases the gravitational field either monotonically decreases becoming nonvanishing in the centre (Hernquist, NFW) or infinite there (Jaffe). In Sect. 5 we analyze the Jaffe and Hernquist mass models for {\it real} spheroidal elliptical galaxies and bulges. We find that, when the Jaffe and Hernquist models are used to describe the stellar population of spheroidal elliptical galaxies and bulges, the presence of the central supermassive black hole avoids the pathological behaviour in the neighbourhood of the centre. But we show that nonetheless both models can {\it only be trusted for} $r \gtrsim 0.2 \, R_e$ ($R_e$ being the effective radius). In Sect. 6 we discuss the NFW model, both for the distribution of galaxies in regular clusters and its application to  pure dark matter halos. In either case, the NFW model loses its validity towards the centre, thereby failing to predict a central cusp both in the distribution of galaxies in regular clusters and in pure dark matter halos. And an application to dark matter annihilation into photons is ruled out. Finally, in Sect. 7 we draw our conclusions.  

\section{Discussion of the statement} 

Before committing ourselves with any specific model described by a spherically symmetric density profile $\rho (r)$, we stress a statement which ensures that {\it a given galactic model} makes sense close to the centre. 

\

\noindent {\bf STATEMENT:} Suppose that an arbitrary spherically symmetric galactic model defined by the density profile $\rho (r)$ has integrated mass profile
\begin{equation}
M (r) \equiv  4 \pi \int_0^r d r^{\prime} ~ r^{\prime 2} \, \rho (r^{\prime}) 
\label{a2}
\end{equation}
such that
\begin{equation}
\label{a1a}
\lim_{r \rightarrow 0} M (r) = 0~. 
\end{equation}
Then, in order for the mass model in question to be {\it physically consistent} near the centre the following two conditions should be met 
\begin{equation}
\label{a1b}
\lim_{r \rightarrow 0} v_c^2 (r) = 0~,
\end{equation}
\begin{equation}
\label{a1c}
\lim_{r \rightarrow 0} g (r) = 0~,
\end{equation}
where $v_c (r)$ denotes the circular velocity while $g (r)$ stands for the gravitational field. 
Indeed, given the fact that in the centre the mass is zero, a vanishing mass must give rise to a vanishing circular velocity and a vanishing gravitational field in the centre. So, whenever condition (\ref{a1a}) is met but one of the conditions (\ref{a1b}) or (\ref{a1c}) is not, then the considered mass model loses its physical meaning in the neighbourhood of the centre.  

\section{A five-dimensional class of spherically symmetric galactic models}

Our models are defined by the following mass density profile
\begin{equation}
\rho ( r ) = \rho_0 \left(\frac{r}{a} \right)^{- \gamma} \left[ 1 + \left(\frac{r}{a} \right)^{\alpha} \right]^{(\gamma - \beta)/\alpha}~,
\label{a1}
\end{equation}
where $\rho_0$, $a$ are arbitrary positive constants, and $\alpha$, $\beta$, $\gamma$ are arbitrary parameters. Models of this sort are mentioned but not thoroughly discussed 
by Mo, van den Bosch and White (\citealt{whitebook}, see also \citealt{binneytremaine}).  

We shall see that for some of them $M (r)$ behaves as $M (r) \to \infty$ for $r \to \infty$, but this fact does not bother us, since realistic astronomical systems are spatially bounded with radius ${\cal R}$, hence the considered mass models should be cut at $r = {\cal R}$. Of course, such a truncation can affect other properties of the models, like for instance isothermality in the case of the regular isothermal sphere~\citep{binneytremaine}. In addition, we shall encounter models which exhibit a central density profile $\rho = {\rm constant}$ which is called a {\it central core}, whereas other models display a {\it central cusp}, namely they have $\rho (r) \to \infty$ as $r \to 0$. {\it A priori}, nobody worries about a central cusp since an infinite central density is not against any physical principle: indeed, the density is merely a derived quantity which cannot be directly measured, and what matters are the integrated mass profile, the circular velocity and the gravitational field: only $v_c (r)$ and $g (r)$ are directly measurable quantities. 

Actually, the main point behind the present analysis is that -- given a certain density profile $\rho (r)$ -- it cannot absolutely be taken for granted that the observable quantities $v_c (r)$ and $g (r)$ possess a physically sensible behaviour towards the centre. Surprisingly, even though several properties of some models included in the considered family have been carefully analyzed, close to the centre so far insufficient attention has been paid to the circular velocity and no attention whatsoever to the gravitational field (with the exception of the regular isothermal sphere,~\citealt{binneytremaine}). 

We should mention that after this Paper was nearly finished we have become aware of the exhaustive analysis of the same class of models described by Eq. (\ref{a1}) carried out in 1996 by Zhao~\citep{zhao}. Nevertheless, the overlap between the two Papers is nearly vanishing, since also Zhao does not consider the behaviour of the gravitational {\it field} $g (r)$. As far as the notations are concerned, the reader can recover Zhao's counterpart of our Eq. (\ref{a1}) by the replacements $\rho_0 \to C$, $r/a \to r$ and $\alpha \to 1/\alpha$.

\section{Mathematical discussion} 

Starting from Eq. (\ref{a1}), the integra\-ted mass profile  
reads
\begin{equation}
M (r) = 4 \pi \rho_0 \, a^3 \int_0^{r/a} d t ~ t^{(2 - \gamma)} \,  
\bigl(1 + t^{\alpha} \bigr)^{(\gamma - \beta)/\alpha}~,
\label{a3}
\end{equation}
whose explicit form is
\begin{eqnarray}
\label{a4}
&\displaystyle M (r) = \frac{4 a^3 \pi \rho_0}{3 - \gamma} \left(\frac{r}{a} \right)^{3 - \gamma} \times \nonumber \\
&\displaystyle \times \, {_2 F}_1 \left[\frac{3 - \gamma}{\alpha}, \frac{\beta - \gamma}{\alpha}; 1+ \frac{3 - \gamma}{\alpha}; - \left(\frac{r}{a}\right)^{\alpha} \right]~,
\end{eqnarray}
where ${_2 F}_1\left(\cdot, \cdot; \cdot; \cdot \right)$ is the confluent hypergeometric function of the second kind. Correspondingly, the circular velocity and the gravitational field are defined as
\begin{equation}
v_c^2 ( r ) \equiv \left(\frac{G}{r} \right) M ( r )~, 
\label{a5}
\end{equation}
\begin{equation}
g(r) \equiv - \, \left(\frac{G}{r^2} \right) M( r )~, 
\label{a6}
\end{equation}
respectively~\citep{binneytremaine}. So, all we need to know is $M ( r )$. 

Specifically, our task is to explicitly investigate the behaviour of $M ( r )$, $v_c^2 (r)$ and 
$g (r)$ as $r \to 0$ for the above-mentioned models, even though our strategy can straightforwardly be extended to {\it arbitrary values} of $\alpha$, $\beta$ and $\gamma$. 

In view of the forthcoming analysis it is therefore instrumental to evaluate $M(r)$, $v_c^2(r)$ and $g(r)$ as $r \to 0$ for $\alpha$, $\beta$ and $\gamma$ in specific ranges. We start with the case $\alpha = 1$, 
$3 \leq \beta \leq 4$ and $0 \leq \gamma < 3$. Correspondingly, we find
\begin{equation}
\lim_{r \rightarrow 0} M(r) = 0~, 
\label{Mgen1}
\end{equation}
while
\begin{equation} 
\lim_{r \rightarrow 0} v_c^2(r) = \begin{cases}
0~, &  0 \leq \gamma < 2~, \\[8pt]
4 \pi G \rho_0 \, a^2~, &  \gamma = 2~, \\[8pt]
\infty~, &  2 < \gamma <3~,
\end{cases} 
\label{v2gen1}
\end{equation}
and
\begin{equation}
\lim_{r \rightarrow 0} g(r) = \begin{cases}
0~, &  0 \leq \gamma <1~, \\[8pt]
- 2 \pi G \rho_0 \, a~, &  \gamma = 1~, \\[8pt]
- \infty~, &  1 < \gamma <3~,
\end{cases} 
\label{Ggen1}
\end{equation}
for any value of $\beta$ in the above range. Next, we address the case $\alpha =2$, $2 \leq \beta \leq 5$ and $\gamma = 0$. Accordingly, we obtain
\begin{equation}
\label{M1a}
\lim_{r \rightarrow 0} M (r) = 0~, 
\end{equation}
\begin{equation}
\label{M1b}
\lim_{r \rightarrow 0} v_c^2 (r) = 0~,
\end{equation}
\begin{equation}
\label{M1c}
\lim_{r \rightarrow 0} g (r) = 0~.
\end{equation}
regardless of the values of $\beta$ in the specified range. As a consequence, in the present case conditions (\ref{a1a}), (\ref{a1b}) and (\ref{a1c}) happen to be automatically satisfied. 

Finally, we proceed to apply these results to the previously considered models.

\begin{enumerate}

\item {\bf NFW model} -- It corresponds to $\alpha=1$, $\beta=3$, $\gamma=1$. The integrated mass profile is
\begin{equation}
M(r) = 4 \pi \rho_0 \, a^3 \left[ {\rm ln} \left(1 + \frac{r}{a} \right)-\frac{r}{r+a}\right]~,
\label{NFW1}
\end{equation}
which meets condition (\ref{a1a}). Owing to its importance, it deserves a thorough discussion, which will be presented in Sect. 6.

\item {\bf Dehnen models} -- They correspond to $\alpha=1$, $\beta=4$, $\gamma < 3$. The integrated mass profile is
\begin{equation}
M(r) = \frac{4 \pi \rho_0 a^3}{3 - \gamma} \left(\frac{r}{r + a} \right)^{3-\gamma}~.
\label{den1}
\end{equation}
Thanks to Eqs. (\ref{Mgen1}), (\ref{v2gen1}) and (\ref{Ggen1}), we see that conditions (\ref{a1a}) and (\ref{a1b}) are obeyed for $0 \leq \gamma < 2$ but conditions (\ref{a1c}) is satisfied for 
$0 \leq \gamma < 1$. So, only for $0 \leq \gamma < 1$ are the Dehnen models physically consistent near the centre.

\item {\bf Hernquist model} -- It is the particular case of the Dehnen models with 
$\gamma=1$. Hence, conditions (\ref{a1a}) and (\ref{a1b}) are met but condition (\ref{a1c}) is not, and we have $g (r) \to - 2 \pi G \rho_0 \, a$ as $r \to 0$. As a consequence, the Hernquist  model is physically inconsistent in the neighbourhood of the centre.

\item {\bf Jaffe model} -- It is a particular case of the Dehnen models with $\gamma=2$.  Thus, only condition (\ref{a1a}) is obeyed but conditions (\ref{a1b}) and (\ref{a1c}) are not, and we have $v_c^2(r) \to 4 \pi G \rho_0 \, a^2$ as $r \to 0$ and $g (r) \to - \, \infty$ as $r \to 0$. Therefore, the Jaffe model is physically inconsistent towards the centre.

\item {\bf Pseudo-isothermal sphere} -- It corresponds to $\alpha=2$, $\beta=2$, $\gamma=0$. The integrated mass profile is
\begin{equation}
M(r) = 4 \pi \rho_0 \, a^2 \left[ r - \, a \, {\rm tan}^{- 1} \left(\frac{r}{a}\right)\right]~.
\label{iso1}
\end{equation}
Owing to Eqs. (\ref{M1a}), (\ref{M1b}) and (\ref{M1c}), conditions (\ref{a1a}), (\ref{a1b}) and (\ref{a1c}) are met. Consequently, the Pseudo-isothermal sphere is physically consistent close to the centre.

\item {\bf Modified Hubble profile} -- It corresponds to $\alpha=2$, $\beta=3$, $\gamma=0$. The integrated mass profile is
\begin{eqnarray}
&\displaystyle M(r) = \frac{4 \pi \rho_0 \, a^3}{r^2+a^2}  -  \\  \label{Hub1}
&\displaystyle - \Bigg[ (r^2 + a^2) \, {\rm sinh}^{- 1} \left(\frac{r}{a} \right) - a \, r \left(1+ \frac{r^2}{a^2} \right)^{1/2} \Bigg]~.  \nonumber 
\end{eqnarray} 
Due to Eqs. (\ref{M1a}), (\ref{M1b}) and (\ref{M1c}), conditions (\ref{a1a}), (\ref{a1b}) and (\ref{a1c}) are satisfied. Hence, the modified Hubble profile is physically consistent near the centre.

\item {\bf Perfect sphere model} -- It corresponds to $\alpha=2$, $\beta=4$, $\gamma=0$. The integrated mass profile is
\begin{eqnarray}
&\displaystyle M(r) = \frac{2 \pi \rho_0 \, a^3}{r^2+a^2} \times   \\  \label{Perf1}
&\displaystyle \times \left[ (r^2+a^2) \, {\rm tan}^{- 1} \left(\frac{r}{a}\right) \nonumber
- a \, r \right]~.
\end{eqnarray}
Thanks to Eqs. (\ref{M1a}), (\ref{M1b}) and (\ref{M1c}), conditions (\ref{a1a}), (\ref{a1b}) and (\ref{a1c}) are met. So, the perfect sphere model is physically consistent in the neighbourhood of the centre.

\item {\bf Plummer sphere model} -- It corresponds to $\alpha=2$, $\beta=5$, $\gamma=0$. The integrated mass profile is
\begin{equation}
M(r) = \frac{4}{3}\pi \rho_0 r^3 \left(1+ \frac{r^2}{a^2} \right)^{-3/2}~.
\label{Plummer1}
\end{equation}
On account of  Eqs. (\ref{M1a}), (\ref{M1b}) and (\ref{M1c}), conditions (\ref{a1a}), (\ref{a1b}) and (\ref{a1c}) are obeyed. So, the Plummer sphere model is physically consistent towards  the centre.

\end{enumerate}

\section{Real spheroidal ellipticals and bulges} 

The analysis carried out so far is formal in nature, since it merely refers to specific abstract models. For instance, models describing the stellar distribution inside spheroidal elliptical galaxies and bulges are invalid close to the centre because of the presence of a supermassive black hole (SMBH). Nonetheless, our previous results are important because they are alarm bells that some models can be pathological also {\it beyond} the SMBH. Below, we will carefully analyze the behaviour of such models in their realistic context.

\

The Dehnen models -- and in particular the Jaffe and Hernquist ones -- have routinely been used to represent the stellar distribution within spheroidal elliptical galaxies and bulges (just to quote a few papers, see~\citealt{hui1995,rix1997,gerhard1998,saglia2000,cappellari2006}). 

Let us therefore discuss the effect of the central SMBH on a generic Dehnen model. Here, the relevant quantity is the {\it dynamical radius} ${\cal R}_g$, where the gravitational field of the SMBH and of the host galaxy are equal~\citep{binneytremaine}. We neglect the dark matter, because the central region of ellipticals and bulges is believed to be baryon dominated. It is then trivial to find that 
${\cal R}_g$ is given by 
\begin{equation} 
{\cal R}_g = a \left\{\left[\frac{4 \pi \rho_0 \, a^3}{(3 - \gamma)M_{\rm SMBH}}\right]^{\frac{1}{3-\gamma}} - \,1 \right\}^{-1}~,
\label{rg} 
\end{equation}
but since the term inside the square brackets is obviously much larger than 1 -- defining $M_e \equiv \bigl(4 \pi \rho_0 \, a^3 \bigr)/3$ -- Eq. (\ref{rg}) boils down to the following approximate expression
\begin{equation}
{\cal R}_g \simeq a \left(1-\frac{\gamma}{3} \right)^{\frac{1}{3-\gamma}} \left( \frac{M_{\rm SMBH}}{M_e}\right)^{\frac{1}{3 - \gamma}}~.
\label{rgapprox} 
\end{equation}
Several relationships exist in the literature between $M_{\rm SMBH}$ and the central 
one-dimensional velocity dispersion of the host galaxy $\sigma (0)$ as evaluated within a given aperture (this point has been carefully discussed in~\citealt{tremaine2002}). In order to be specific -- choosing the aperture $R_e/8$ -- $M_{\rm SMBH}$ reads~\citep{merrittferrarese2001}
\begin{equation}
M_{\rm SMBH} \simeq 1.30 \cdot 10^8 \left(\frac{\sigma (0)}{200 \, {\rm km} \, {\rm s}^{- 1}} \right)^{4.72} \, M_{\odot}~.
\label{24022020a} 
\end{equation}
Thus, we conclude that the Dehnen models can make sense for a galactocentric distance {\it larger} than ${\cal R}_g$ as provided by Eqs. (\ref{rg}) or (\ref{rgapprox}) (more about this, later).

\bigskip

\centerline{ \ \ \  * \ \ \  * \ \ \  * \ \ \ }

\medskip 

As a next step, we focus our attention on the Jaffe and Hernquist models. Since $a = 0.55 \, R_e$ for the Hernquist model, and  $a = 1.31 \, R_e$ for the Jaffe one, by specializing Eq. (\ref{rgapprox}) to these cases, we get
\begin{equation}
{\cal R}_{g, J} \simeq 0.44 \, R_e \left(\frac{M_{\rm SMBH}}{M_e}\right)~,
\label{rgapproxJ} 
\end{equation}
and
\begin{equation}
{\cal R}_{g, H} \simeq 0.45 \, R_e \left(\frac{M_{\rm SMBH}}{M_e}\right)^{1/2}~.
\label{rgapproxH} 
\end{equation}
So, only for galactocentric distances larger than either ${\cal R}_{g, J}$ or ${\cal R}_{g, H}$ 
can the Jaffe or the Hernquist models be regarded  as a realistic description of the stellar population of spheroidal ellipticals and bulges. 

Incidentally, a slightly different discussion of the Hernquist model is contained in (\citealt{binneytremaine}, see Fig. 4.20), where -- denoting by $M_g$ the mass of the galaxy -- for $M_{\rm SMBH} = 0.002 \, M_g$ and $M_{\rm SMBH} = 0.004 \, M_g$ it is found 
${\cal R}_{g, H} \simeq 0.026 \, R_e$ and ${\cal R}_{g, H} \simeq 0.037 \, R_e$, respectively.

\bigskip

\centerline{ \ \ \  * \ \ \  * \ \ \  * \ \ \ }

\medskip 

We prefer to work henceforth with the dimensionless quantities defined as follows. 

\begin{enumerate} 

\item Radial distance: $r/R_e$.

\item Mass density: $\rho/\rho_0$.

\item Integrated mass profile: $M (r)/ \bigl(4 \pi \, \rho_0 \, R_e^3 \bigr)$. 

\item Square circular velocity: $v_c^2 (r)/\bigl(4 \pi G \, \rho_0 \, R_e^2 \bigr)$. 

\item Gravitational field: $g (r)/\bigl(2 \pi G \, \rho_0 \, R_e \bigr)$. 

\end{enumerate} 
We will replace $R_e$ by $a_{\rm NFW}$ for the NFW model.

\

\begin{figure}
\centering
\includegraphics[width=.48\textwidth]{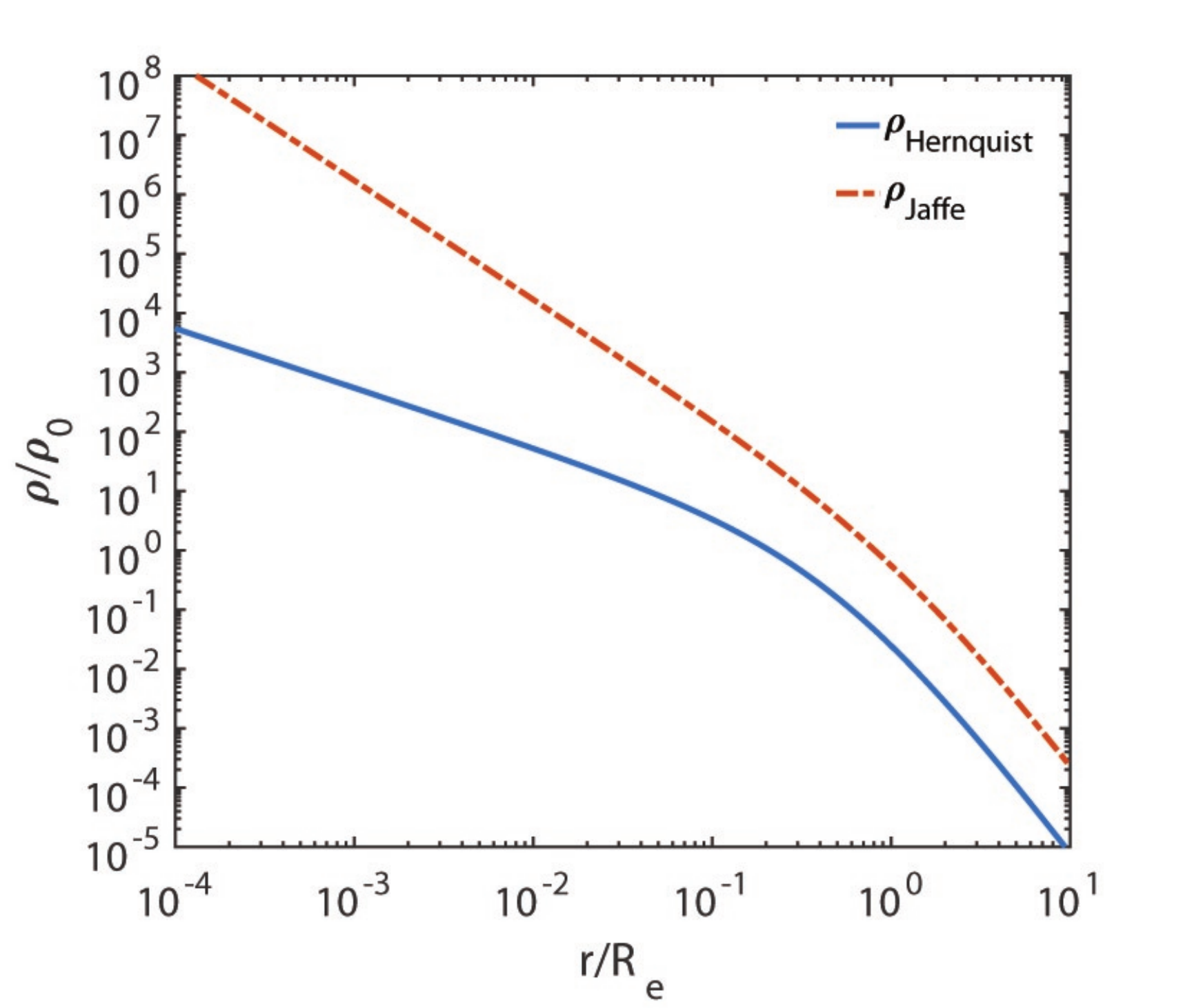}
\caption{\label{figC1} We report on the vertical axis $\rho/\rho_0$ and on the horizontal axis $r/R_e$, both in logarithmic scale.}
\end{figure}

\begin{figure}
\centering
\includegraphics[width=.48\textwidth]{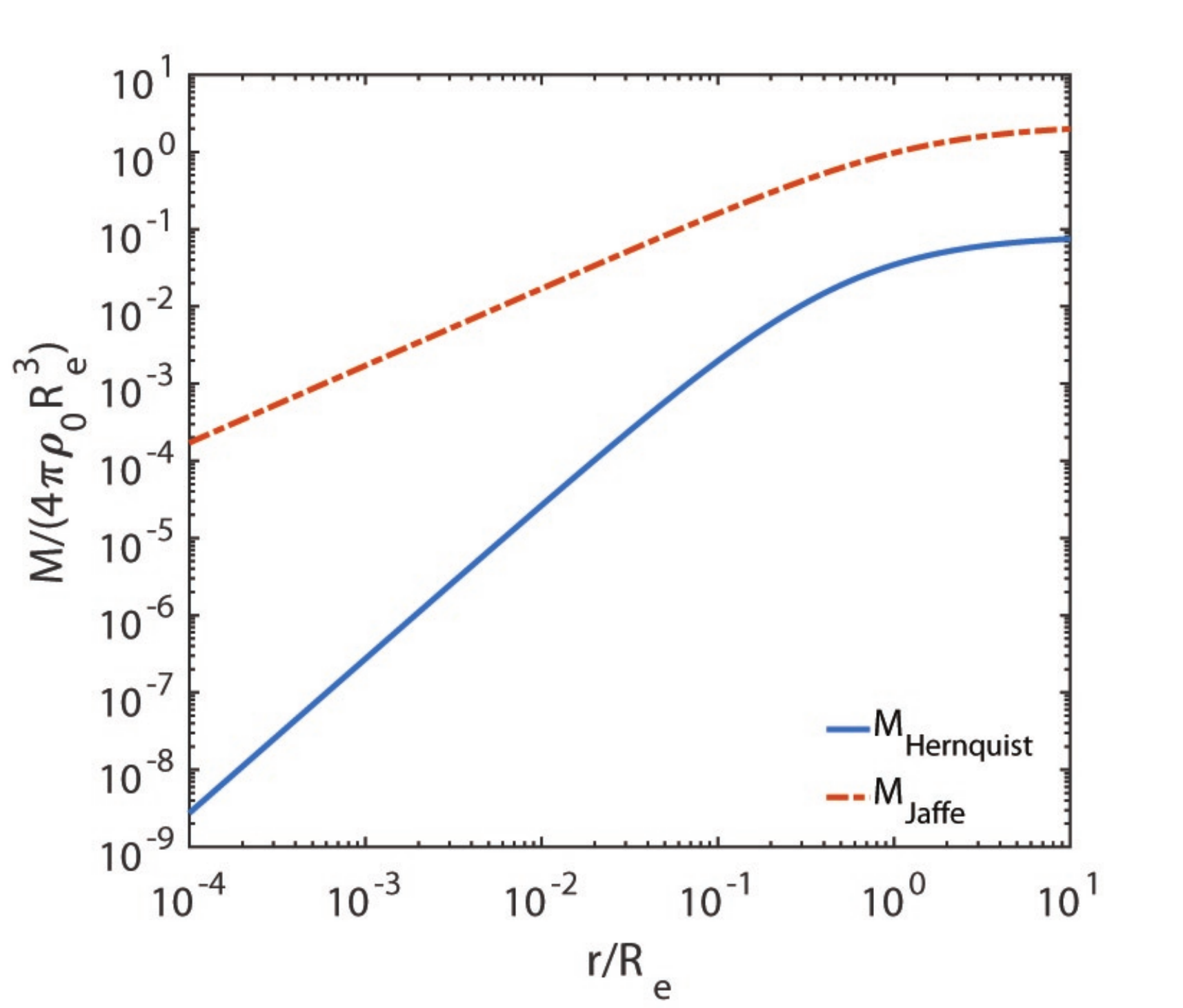}
\caption{\label{figC2} We show on the vertical axis $M (r)/\bigl(4 \pi \, \rho_0 \, R_e^3 \bigr)$ and $r/R_e$ on the horizontal axis, both in logarithmic scale.}
\end{figure}

\begin{figure}
\centering
\includegraphics[width=.48\textwidth]{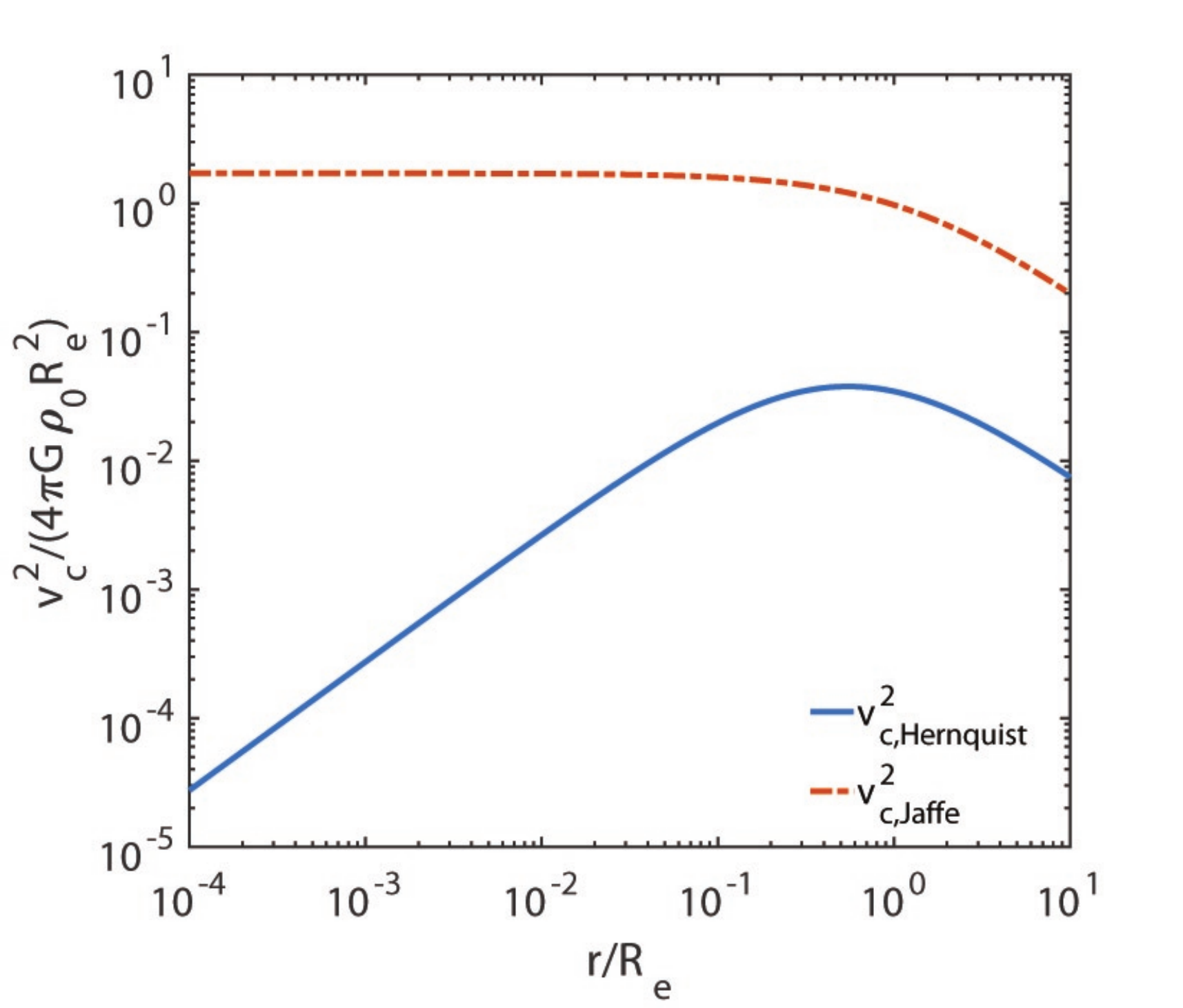}
\caption{\label{figC3} We exhibit on the vertical axis $v_c^2 (r)/\bigl(4 \pi G \, \rho_0 \, R_e^2 \bigr)$  and $r/R_e$ on the horizontal axis, both in logarithmic scale.}
\end{figure}

\begin{figure}
\centering
\includegraphics[width=.48\textwidth]{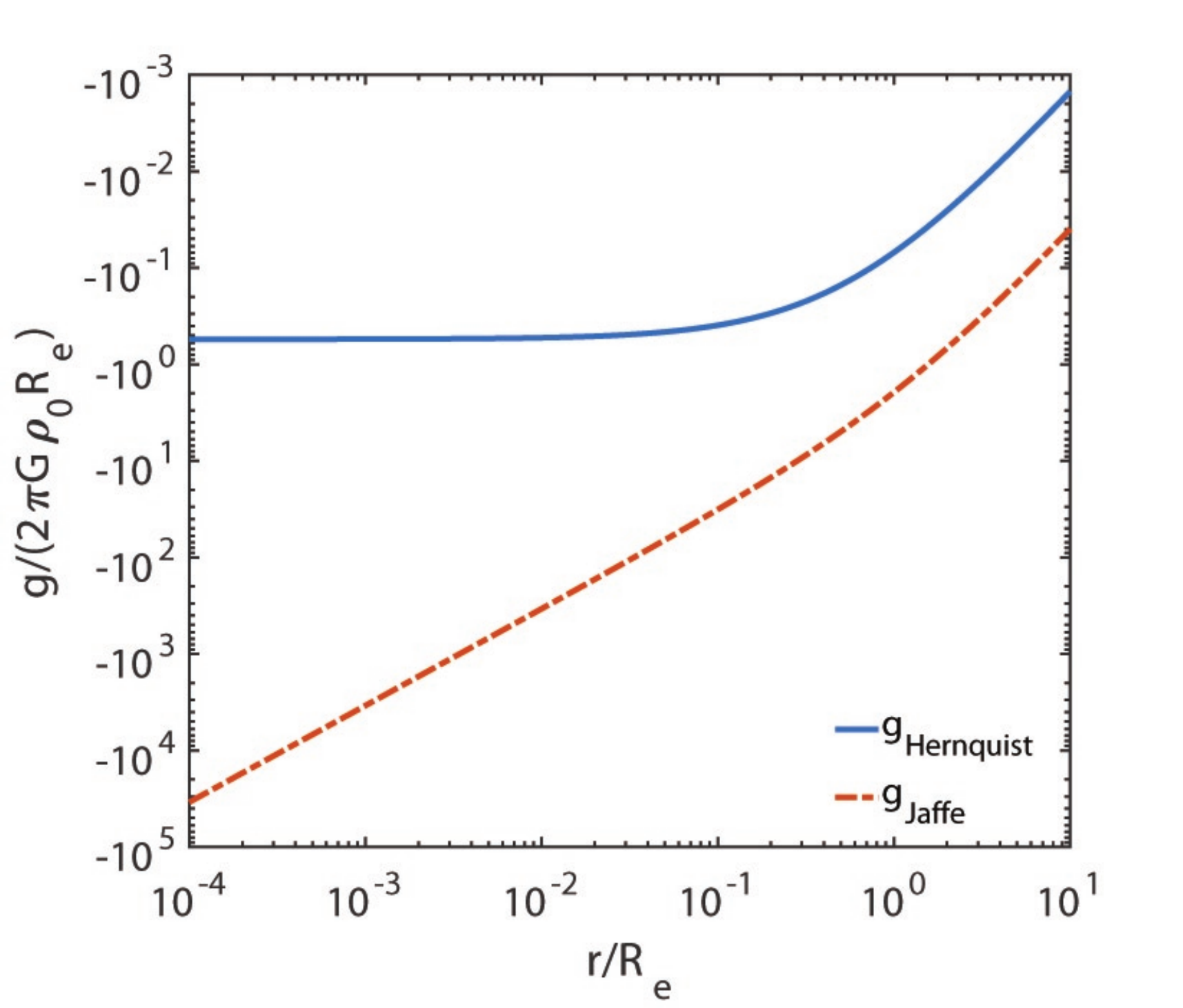}
\caption{\label{figC4} We report on the vertical axis $g (r)/\bigl(2 \pi G \, \rho_0 \, R_e \bigr)$ and $r/R_e$ on the horizontal axis, both in logarithmic scale.} 
\end{figure}

We are now in a position to assess the validity of the Jaffe and Hernquist models. Because we are interested to investigate in great detail what happens around the centre, we plot 
$\rho/\rho_0$, $M (r)/ \bigl(4 \pi \, \rho_0 \, R_e^3 \bigr)$, $v_c^2 (r)/\bigl(4 \pi G \, \rho_0 \, R_e^2 \bigr)$ and $g (r)/\bigl(2 \pi G \, \rho_0 \, R_e \bigr)$ versus $r/R_e$ in logarithmic scales in Figs.~\ref{figC1},~\ref{figC2},~\ref{figC3} and~\ref{figC4}. 

\begin{figure}
\centering
\includegraphics[width=.48\textwidth]{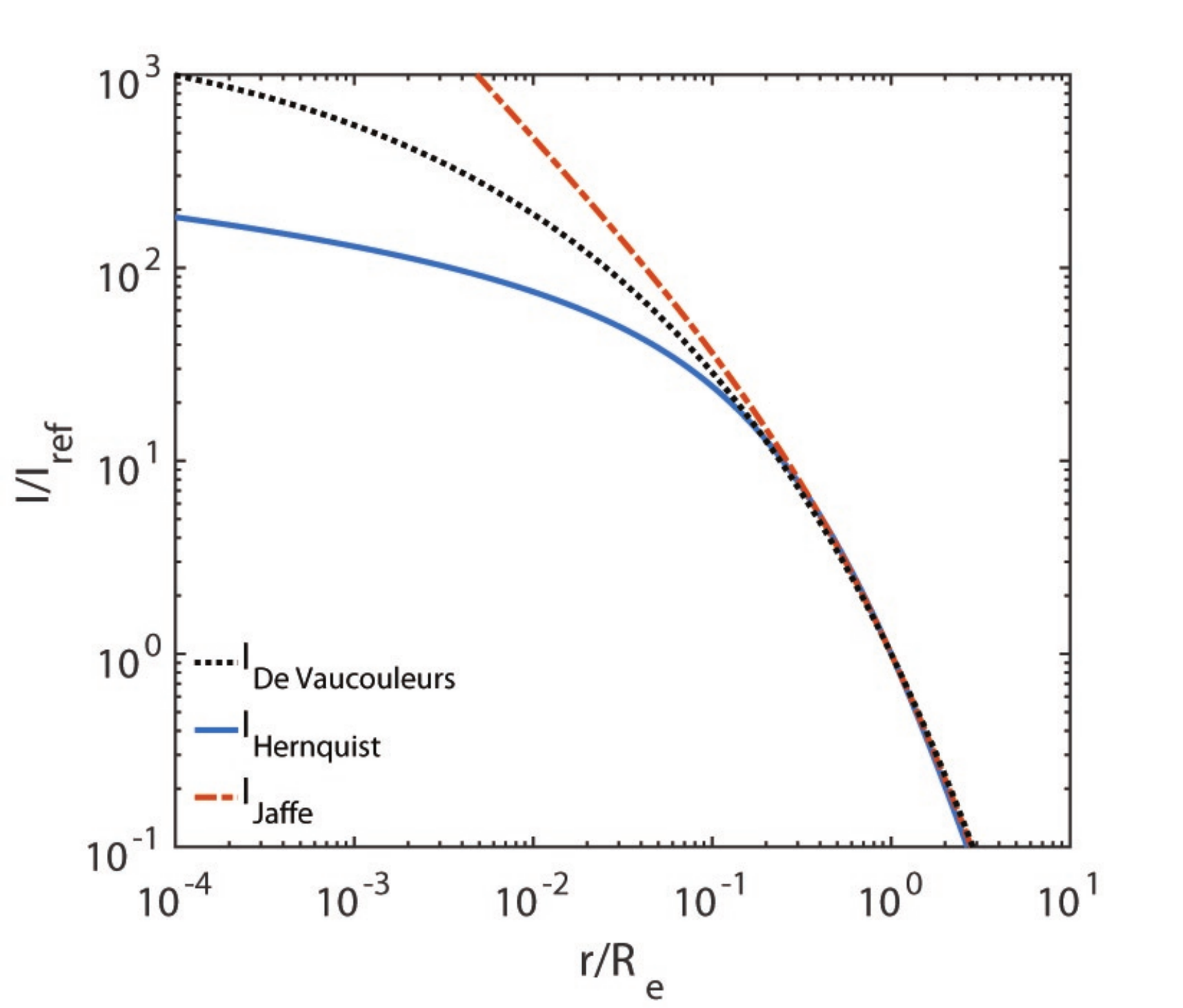}
\caption{\label{fig2306a} We show the projected Jaffe and Hernquist models as well as the De Vaucouleurs law, both in logarithmic scale. In all cases ${\rm I}/{\rm I}_{\rm ref}$ is the dimensionless surface brightness.}
\end{figure}

The departure from similarity of the two models takes place around $r \simeq 0.2 \, R_e$, where it starts to become larger and larger as the galactocentric distance gets smaller and smaller.  Moreover, the circular velocity curve for the Hernquist model is physically very well behaved while for the Jaffe model it is not. In addition, the gravitational field does not show any turn towards 0 for both models. Fortunately, we can make sense out of such a behaviour by recalling that historically both models have been devised in order to reproduce the De Vaucouleurs surface brightness profile upon projection, assuming a constant mass-to-light ratio. Accordingly, their shape should nearly coincide at, say, $r = 2 \, R_e$, as indeed it takes place in the considered figures. We are thus led to the guess that both models fail to fit the De Vaucouleurs law in projection for $r \lesssim 0.2 \, R_e$.  A check of our guess can be obtained by projecting these models onto the sky. The results are shown in Fig.~\ref{fig2306a}.

We see that Fig.~\ref{fig2306a} beautifully shows that indeed both the Jaffe and Hernquist model {\it can only be trusted} for $r \gtrsim 0.2 \, R_e$, if we want to stick to a {\it constant} luminous mass-to-light ratio $\Upsilon_{\rm lum}$ for ${\cal R}_g < r \lesssim 0.2 \, R_e$ (our conversion from surface mass density to surface brightness has been performed by assuming $\Upsilon_{\rm lum} = {\rm constant}$). Taking these model seriously in the latter range an unphysical gradient in $\Upsilon_{\rm lum}$ would necessarily show up, which could be confused with a colour gradient or a gradient of the total mass-to-light ratio $\Upsilon_{\rm tot}$, which might erroneously be interpreted as evidence for dark matter.

\section{NFW model} 

\begin{figure}
\centering
\includegraphics[width=.48\textwidth]{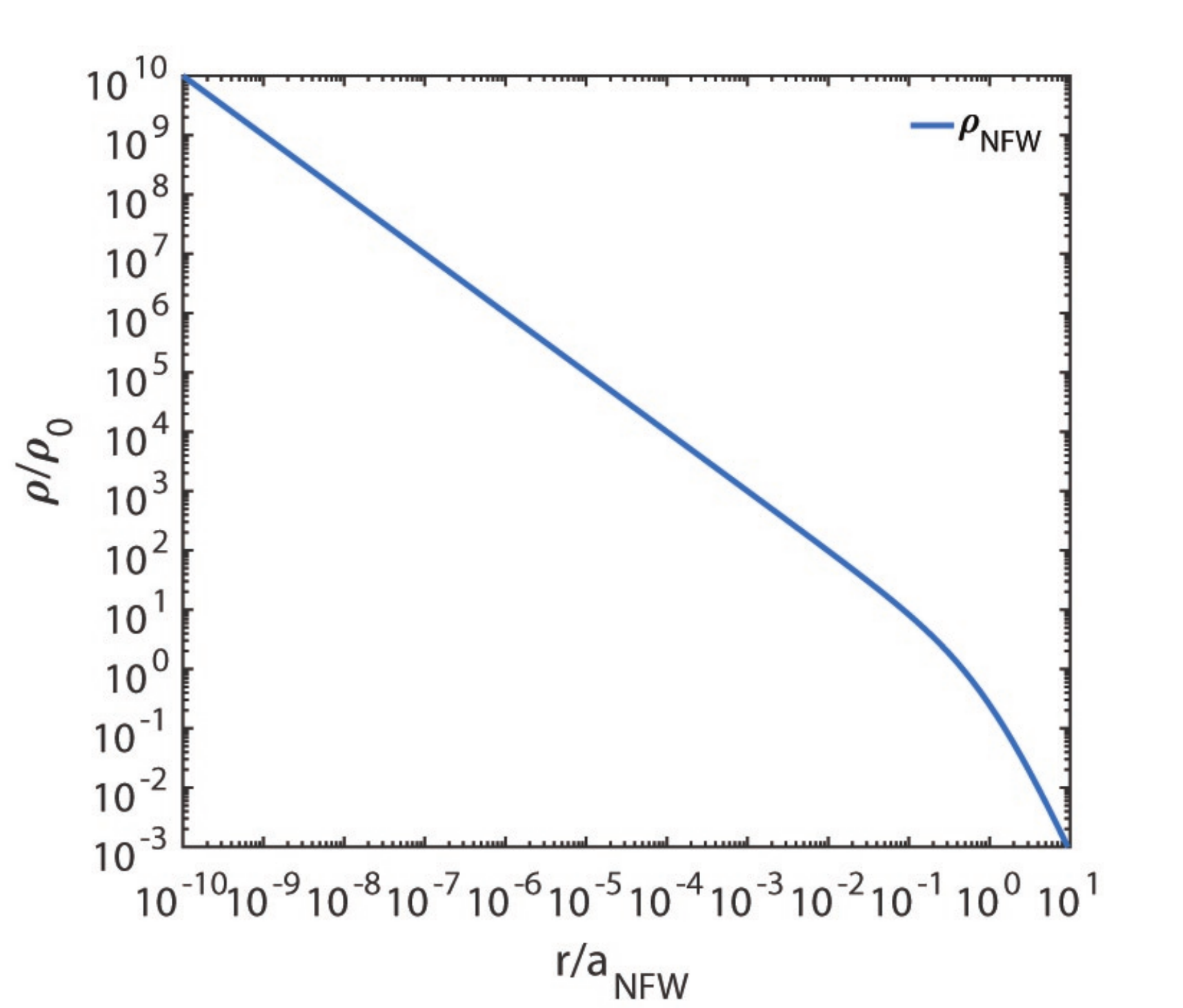}
\caption{\label{fig7a2} We exhibit on the vertical axis $\rho (r)/\rho_0$ and on the horizontal axis $r/a_{\rm NFW}$, both in logarithmic scale.}
\end{figure}

\begin{figure}
\centering
\includegraphics[width=.48\textwidth]{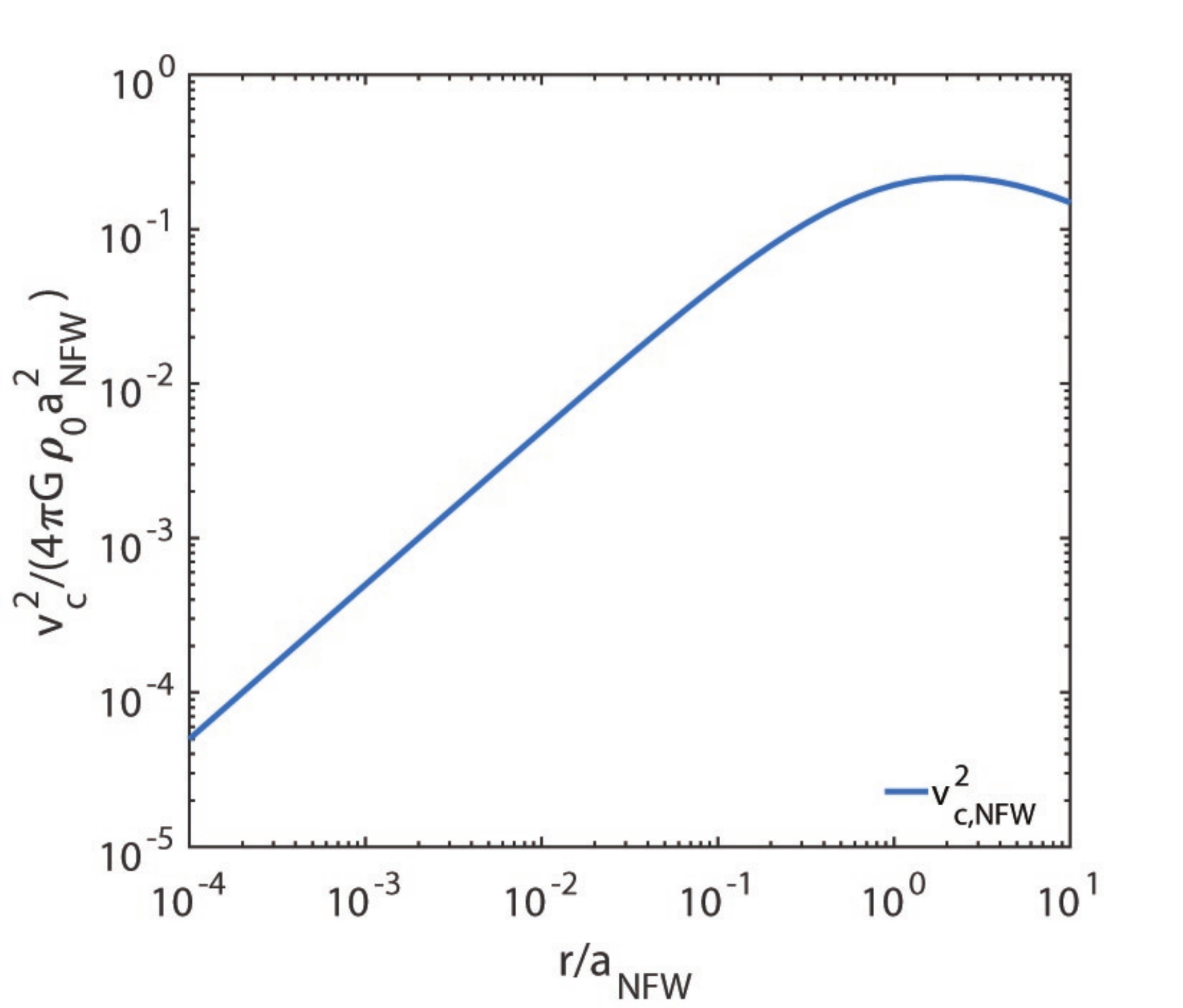}
\caption{\label{fig3} We report on the vertical axis $v_c^2 (r)/\bigl(4 \pi G \, \rho_0 \, a^2_{\rm NFW} \bigr)$ and on the horizontal axis $r/a_{\rm NFW}$, both in logarithmic scale.}
\end{figure}

\begin{figure}
\centering
\includegraphics[width=.48\textwidth]{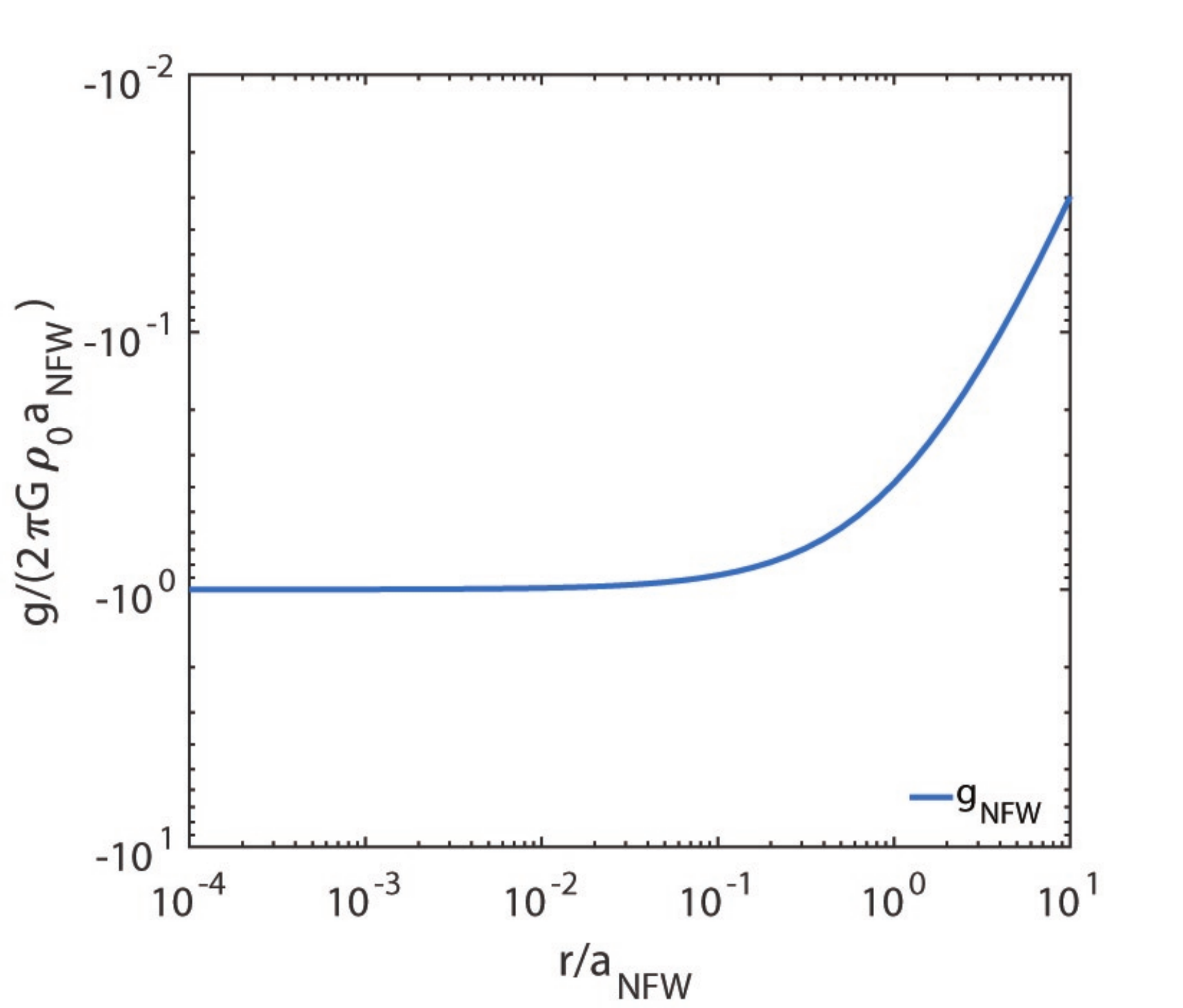}
\caption{\label{fig4} We show on the vertical axis $g (r)/\bigl(2 \pi G \, \rho_0 \, a_{\rm NFW} \bigr)$ and on the horizontal axis $r/a_{\rm NFW}$, both in logarithmic scale.}
\end{figure}

Let us come back to the NFW profile, whose explicit form is 
\begin{equation}
\rho ( r ) = \rho_0 \left(\frac{r}{a_{\rm NFW}} \right)^{- 1} \left[ 1 + \left(\frac{r}{a_{\rm NFW}}  \right) \right]^{- 2}~, 
\label{24022010c}
\end{equation}
which we plot in a log-log diagram in Fig.~\ref{fig7a2}. Moreover, we get the dimensionless square circular velocity $v^2_c (r)/\bigl(4 \pi G \, \rho_0 \, a^2_{\rm NFW} \bigr)$ and the dimensionless gravitational field $g (r)/\bigl(2 \pi G \, \rho_0 \, a_{\rm NFW} \bigr)$ upon the replacement $a \to a_{\rm NFW}$ in Eq. (\ref{NFW1}) and by employing Eqs. (\ref{a5}) and (\ref{a6}), respectively. These quantities are plotted versus $r/a_{\rm NFW}$ in Figs. (\ref{fig3}) and (\ref{fig4}). 

While the behaviour of $v_c^2 (r)$ shown in Fig.~\ref{fig3} looks physical and in agreement with our statement, the behaviour of $g (r)$ exhibited in Fig.~\ref{fig4} implies that this is not the case. Hence, the NFW model fails close enough to the centre. We stress that the situation is presently worse as compared to the one discussed in Sect. 5, since we have no handle to tell at which distance from the centre the NFW model breaks down. 

\subsection{NFW model and regular galaxy clusters} 

Nowadays, the overall distribution of galaxies in regular clusters is believed to be well described by an NFW model with $a_{\rm NFW} = R_{\rm vir}/c_{\rm gal}$, where $R_{\rm vir}$ denotes the virial radius and the galaxy concentration $c_{\rm gal}$ ranges from $c_{\rm gal} = 3.7$~\citep{carlsberg1997} to $c_{\rm gal} = 4.2$~\citep{vandermarel2000}. Unfortunately, the galaxy distribution in the central region is more uncertain. According to Adami et al., the luminosity profile of the brightest galaxies is significantly cusped in the center of the clusters (regardless of the  redshift), but the luminosity profile of the fainter galaxies is significantly better fitted by a cored model~\citep{adami2001}. But Lin et al. claim that {\it all} galaxies are distributed according to a model (\ref{a1}) with $\alpha = 1.07$, $\beta = 3$, $\gamma = 1$ and $c_{\rm gal} = 2.71$, which is almost undistinguishable from an NFW profile~\citep{lin2004}. But according to our result the galaxy distribution {\it cannot be} represented by an NFW profile all the way down to the centre.

\subsection{NFW model and dark matter halos} 

It is well known that the NFW profile provides the classic analytic fit to the N-body simulations of collisionless cold dark matter particles. Within this context we have $a_{\rm NFW} \equiv r_{200} /c_h$, where $r_{200}$ is the radius where the overdensity is $200$  times larger than the mean cosmic density -- currently considered as the virial radius -- while $c_h$ is the halo concentration parameter, which depends on both the halo mass and its redshift~(\citealt{whitebook}). But for the present analysis we do not need to commit ourselves with any specific value. 

Thus, to the extent that dark matter halos are correctly described by the NFW model, no central cusp is predicted because the model {\it does not} make physical sense near the centre. From an observational standpoint, the situation is identical to the one encountered in Subsect. 6.1: while at large enough distances from  the centre there is no doubt that the NFW model is physically consistent -- and observationally correct -- we are totally unable to tell at which central distance it breaks down. We nevertheless note that even though this looks disappointing from a conceptual point of view, it makes the NFW model {\it in agreement with observations}. Indeed, the cusp has not been found even where it should, namely in bulgeless galaxies like the low surface brightness and dwarf ones (see e.g.~\citealt{adams2014,oh2015} and references therein).

As far as the dark matter halos are concerned -- even if embedded in real galaxies -- the NFW cusp with mass density profile $\rho (r)$ has been used by many researchers for a very specific purpose. But to appreciate this point some preliminaries are compelling. Let us assume that the non-baryonic cold dark matter is made of standard weakly interacting massive particles (WIMPs), which couple to ordinary ones with the weak interaction strength. Then, their mass $M_{\rm WIMP}$ must obey the lower bound $M_{\rm WIMP} \gtrsim 45 \, {\rm GeV}$, since otherwise they would have been detected at CERN as decay products of the ${\rm Z}^0$ (but some unconventional WIMPs can avoid such a bound, see e.g.~\citealt{escudero2017}). Even if WIMPs must be almost stable in order to survive until the present -- otherwise they could not be the dark matter today -- there is a small but nonvanishing probability that two of them annihilate into standard model particles. This possibility has been recognized long ago by many authors (for a review and a list of references, see~\citealt{jungman1996,bertone2005,berstrom2012}). The crucial point is that -- since we are dealing with a two-body annihilation process -- the photon flux is proportional to the WIMP square density, namely $\rho^2 (r)$.  As a consequence, the presence of the NFW cusp would {\it greatly enhance the resulting photon flux}: this is the key-point. Such a possibility is ruled out.

\section{Conclusions}  

We have first stressed the statement according to which any spherically symmetric galactic model whose integrated mass profile $M (r) \to 0$ as $r \to 0$ is physically consistent in the neighborhood of the centre only provided that the circular velocity $v_c (r) \to 0$ as $r \to 0$, and the gravitational field $g (r) \to 0$ as $r \to 0$. We have next applied the considered statement to some most used models from a class of five-parameter self-gravitating spherical galactic models, which are most frequently used in astrophysics and cosmology, like the Hernquist, Jaffe and NFW ones. 

As is well known, the stellar population of spheroidal elliptical galaxies and of bulges are often described by either the Jaffe model or by the Hernquist one. We have shown that in both cases -- even taking the central SMBH into account -- they can be trusted only for galactocentric distances {\it larger than about 0.2 effective radii}. 

We have next addressed the distribution of galaxies in regular clusters and the density profiles of pure dark matter halos, which are both believed to be well represented by an NFW model, even if with different values of the parameters. We have demonstrated that in either case such a description must break down towards the centre, thereby avoiding the  central cusp which is instead predicted by the NFW model. As far as regular clusters are concerned, the central galactic density profile can be determined by improving photometry, while for pure dark matter halos it is unclear whether N-body simulations with resolution better than about $1 \, {\rm kpc}$ solve the problem {\it before} the baryonic infall. Moreover --  at present -- in both cases we have no idea where such a failure starts to takes place. 


Finally, recently the important possibility that baryons alter the halo density profile close to  the galactic center after their infall -- in such a way to wipe out the dark matter cusp, replacing it with a core -- has started to be systematically investigated by employing models of the form (\ref{a1}) for suitable values of the parameters not considered here (see e. g.~\citealt{dekel2017,freundlich2020}). We plan to extend the present analysis to that very interesting case in a future publication.


\section*{Acknowledgments} 

We warmly thank Magda Arnaboldi, Patrizia Caraveo, Stefano Ettori, Ortwin Gerhard and Andrea Macci\`o for discussions and criticism, and Giancarlo Setti for a careful reading of the manuscript. The work of MR is supported by an INFN TAsP grant and GG acknowledges contribution from the grant ASI-INAF 2015-023-R.1.

\section*{Data availability}
The data underlying this article will be shared on reasonable request to the corresponding author.

\label{lastpage}

\end{document}